\documentclass{epl}

\def\(({\left(}
\def\)){\right)}
\def\[[{\left[}
\def\]]{\right]}

\def\la{\langle}
\def\ra{\rangle}

\newcommand{\be}{\begin{equation}}
\newcommand{\bea}{\begin{eqnarray} \nonumber}
\newcommand{\ee}{\end{equation}}
\newcommand{\eea}{\end{eqnarray}}

\title{On temperature chaos in Ising and XY Spin Glasses}
\author{Florent Krz\c{a}ka{\l}a}
\institute{
Dipartimento di Fisica, INFM and SMC, Universit\`a di Roma {\em La Sapienza} \\
P. A. Moro 2, 00185 Roma, Italy}
\begin{document}
\maketitle
\vspace{-0.5cm}
{\it Il est plus facile de se vieillir que de se rajeunir.} \\
\hspace{4cm}{Jules Renard, Journal}\\

\begin{abstract}
We argue that the chaotic temperature effect predicted in Ising spin glasses should be stronger when one considers continuous (XY, Heisenberg) kind of spins, due to bigger entropic fluctuations. We then discuss the behavior of 3d spin glasses using the Migdal-Kadanoff Renormalization Group for Ising and XY spins, where we show explicitly that the chaotic length scale, the length beyond which equilibrium configurations are completly reshuffled when one changes temperature, could be far smaller for XY than for Ising spins. These results could thus explain why experiments are seeing a stronger rejuvenation effect for continuous spins.
\end{abstract}

One of the most striking features of spin glasses~\cite{Young98} are the phenomena of {\it rejuvenation} and {\it memory}~\cite{JonasonVincent98} when temperature hysteresis cycles are performed. When a spin glass approaches equilibrium, it ages and reduces its susceptibility; however, if one lowers the temperature after one such aging at temperature $T_1$, one sees a restart (a rejuvenation) of the susceptibility aging at temperature $T_2<T_1$, while memory of the previous aging can be retrieved when heating back. More impressive is maybe the fact that many such memory cycles can be performed~\cite{DupuisVincent01}. One of the possible explanations is to consider the  {\it temperature chaos} effect, that is the possibility that equilibrium spin configurations at temperatures $T_1$ and $T_2$ are uncorrelated, a property that has been predicted to hold in spin glasses many years ago~\cite {FisherHuse86,BrayMoore87}. The presence of memory is then compatible with rejuvenation because of length scale separation: dynamics at $T_2<T_1$ is so slow that it is not fast enough to destroy the previous ordering (see for instance the ghost domains of~\cite{YoshinoLemaitre00}). In fact most of spin glass dynamics (see~\cite{BouchaudCugliandolo98} for a classic review) could be explained if one assumes that there is a growing coherence length scale during aging, a strong sensitivity of equilibrium configurations to temperature and a separation of length scales at each temperature~\cite{BerthierViasnoff02}. Another very important experimental result is that rejuvenation is stronger for Heisenberg than for Ising spins~\cite{DupuisVincent01,JonssonYoshino02,JonssonYoshino02b,BertDupuis03}: the sharpness of this effect appears to decrease continuously with the spin anisotropy.

The purpose of this paper is to argue that this might be simply due to a stronger temperature chaos for continuous (XY, Heisenberg) spins, and to provide a simple model to illustrate this assumption.  Also, since it has been explained recently ``why temperature chaos is hard to observe''~\cite{AspelmeierBray02}, we would like to show how it may actually be easier to see. The paper is organized as follows: first, we discuss generally temperature chaos in spin glasses, then we introduce a $3d$ spin glass model interpolating between $XY$ and Ising spins and the Midgal-Kadanoff Renormalization Group we will use to study it. We will present results showing that chaos is stronger when considering XY type of spins, and more generally when there are more degrees of freedom in the system. We will conclude by a general discussion.

\section{Temperature Chaos in Spin Glasses} Let us start by the simple thermodynamic argument of~\cite{BrayMoore87}, and consider a $3d$ Edwards-Anderson~\cite{EdwardsAnderson75} spin glass in the scaling/droplet~\cite{BrayMoore84,FisherHuse86} approach: take two equilibrium states at temperature $T<T_c$ differing by a very large droplet of characteristic size $\ell$. Then the two states have free-energies that differ by $\Delta F(T) = \Delta E - T \Delta S \approx \Upsilon\((T\)) {\ell}^{\theta}$, where $\Upsilon\((T\))$ is the {\it energy stiffness} coefficient. When one changes the temperature by $\delta T$ then $\Delta F(T+\delta T) \approx \Delta E - (T+\delta T) \Delta S$ so that $\Delta F(T+\delta T) \approx \Upsilon\((T\)) {\ell}^{\theta} - \delta T  \Delta S $. In this phenomenological approach, the entropy difference is associated with the droplet's surface so that $\Delta S$ has a random sign and a typical magnitude $\sigma\((T\)) \ell^{d_s/2}$, where $\sigma\((T\))$ is called the {\it entropy stiffness} and $d_s$ is the fractal dimension of the droplet's surface. If ${d_s}/{2}>{\theta}$, which follows from droplet theory, then $\Delta F(T+\delta T)$ can change sign between $T$ and $T+\delta T$ for length scales greater than 
\be
\ell_{c} \propto \(( 
\frac
{\Upsilon\((T\))}{\sigma\((T\)) \delta T}
\))^{1/{\xi}}
\label{lstar}
\ee
with $\xi=d_s/2 - \theta$. So, when temperature is changed, equilibrium configurations are changed on scales greater than $\ell_c$. The crucial point in the argument lies in these large cancellations between $E$ and $S$ in the equilibrium value of $F$: these make the equilibrium state very sensitive to changes in $T$. This picture of chaotic temperature dependence has been largely confirmed within various scaling approaches~\cite{BanavarBray87,NifleHilhorst92,CieplakLi93}, in the naive mean field approximation of the spin glass phase~\cite{KrzakalaMartin02} as well as in a number of toy models~\cite{SalesYoshino02b,KrzakalaMartin02,SasakiMartin02b,Sche}. There are now analytical results for such a dependence in mean field models~\cite{RizzoCrisanti02} but, however, weak enough to be very hard to see in simulations~\cite{BilloireMarinari00}. In the case of finite dimensional Ising spin glasses, it is still not clear if there is (or not) temperature chaos~\cite{BilloireMarinari00,PiccoRicci01,TakayamaHukushima02}, probably because $\ell_c$ is quite big in that case, as claimed by~\cite{AspelmeierBray02}. Thus, temperature chaos does {\it probably} exist in finite dimensional spin glasses but is very hard to see and to confirm so that experimental relevance could be (and has indeed been~\cite{PiccoRicci01,BerthierBouchaud02}) questioned if one has to consider length scales bigger than those that could be reached experimentally. Indeed, other explanations for rejuvenation have been provided~\cite{BerthierBouchaud02,BerthierHoldsworth02,BerthierViasnoff02} and debates and controversies are still open~\cite{BerthierBouchaud03,JonssonYoshino02b}. The experimental fact we are interested in here is, again, that in experiments rejuvenation is found to be stronger for Heisenberg than for Ising spins. 

A simple explanation would thus be that chaos is also stronger for continuous spins. Why should it be so? Looking at eq.(\ref{lstar}), one sees that chaos will increase if energy (exponent or stiffness) decreases and if entropy (exponent or stiffness) increases. This has a simple interpretation: chaos is due to fluctuations of energy and entropy (remember that, in droplet theory, there are more fluctuations in free energies when $\theta$ {\it decreases}); therefore the more fluctuations there are, the more chaos is important! One then may expect that Heisenberg or XY spins, which have many more degrees of freedom, will indeed have more entropy, thus lead to more fluctuations. The purpose of the following study is to illustrate this point in (some) special cases. What we need for studying temperature chaos is a spin glass model which is both (i) tractable enough in order to be able to compute $\ell_c$ and to study the chaotic temperature dependence and (ii) which allows us to study continuous as well as Ising spins.  Following early studies on spin glasses~\cite{SouthernYoung77,BanavarBray87,NifleHilhorst92}, a natural answer is to use the Migdal-Kadanoff Renormalization Group (MKRG)~\cite{Migdal75}.

\section{Migdal-Kadanoff Renormalization Group} 
\begin{figure}
\includegraphics[width=0.42\textwidth]{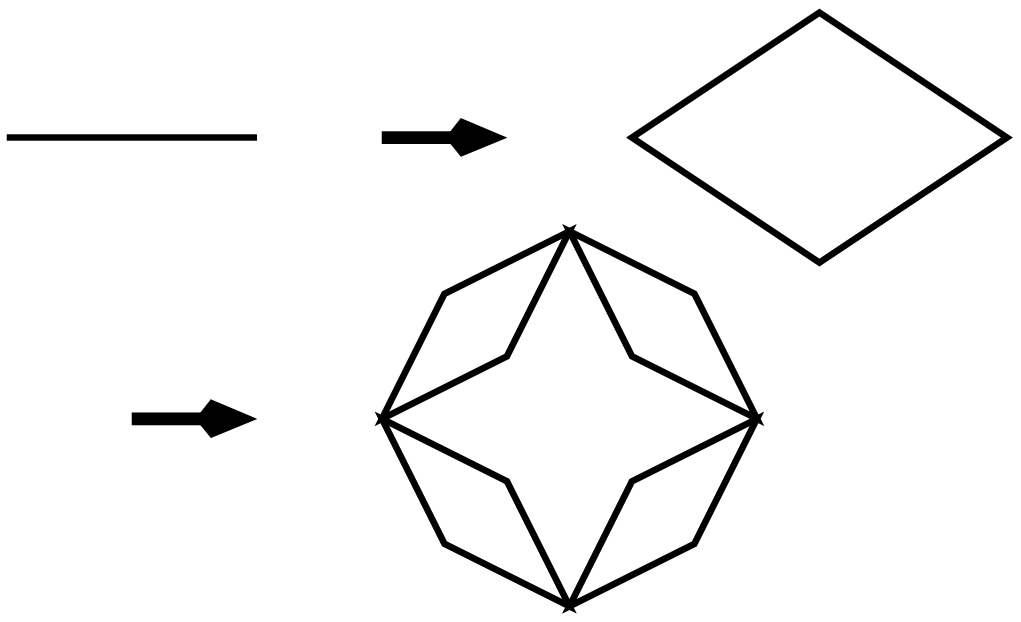}
\includegraphics[width=0.56\textwidth]{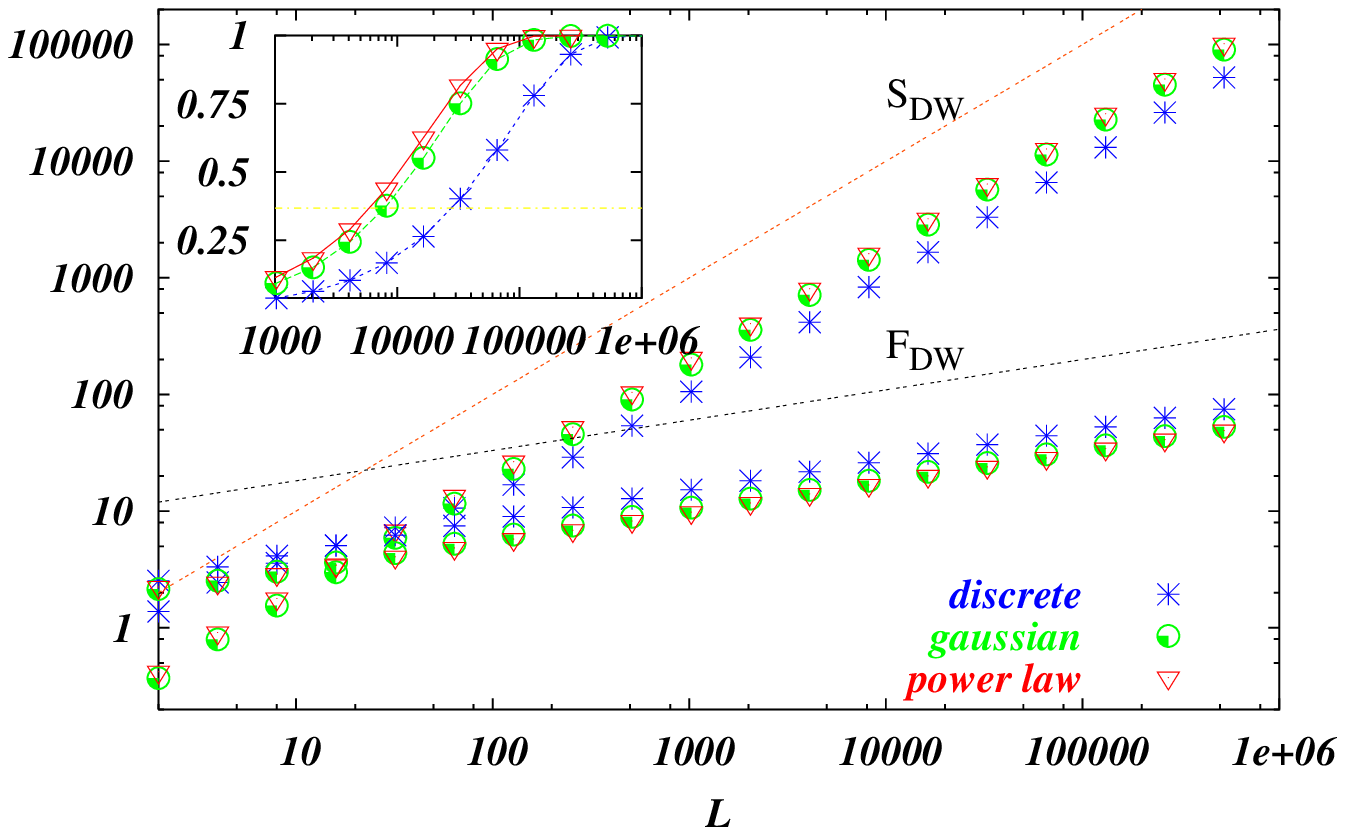}
\caption{Left: Construction of a hierarchical lattice, for dimension $d$ one needs $2^{d-1}$ branches. Right: entropy and energy of an Ising spin glass with  $\pm 1$, Gaussian and power-law tail distributions, rescaled to obtain $T_c=1$; energy decreases and entropy increases while distributions get broader, but exponents (here plotted with $\theta=0.26$ and $d_s/2=1$) remain the same. The inset shows the distance between two replicas at $T=0.1$ and $T=0.11$ versus L: broader distributions lead to more chaos.}
\label{fig_lattice}
\label{fig_Ising}
\end{figure}

It is sometimes convenient to see the MKRG as an exact resolution on a hierarchical lattice (see fig.\ref{fig_lattice}). Since  it is not possible to write directly an exact recursion relation for continuous spins, we will use a {\it q-states} clock model, as introduced by~\cite{CieplakBanavar92} for which the two limits $q=2$ (Ising spins) and $q=\infty$ (XY spins) display a similar droplet-like ordering with a finite $T_c$ as well as the expected chaotic temperature dependence~\cite{CieplakLi93}. We will try here to be more quantitative, following the recent work~\cite{AspelmeierBray02} in the Ising case.
Let us define the $q$-states clock model, following~\cite{CieplakBanavar92}, with a very general random gauge Hamiltonian
\be
H=-\sum_{<ij>}^N J_{ij}\((\phi_i-\phi_j\))
\label{hamiltonian}
\ee
where $J_{ij}(\delta \phi)$ could be any function defined for $\delta \phi$ having values restricted to $2 \pi k/q$, with $k=0,1,2,...,(q-1)$. It is then possible to write a one dimensional decimation procedure, for three consecutive spins a,b,c on a chain~\cite{CieplakBanavar92}
\bea
J_{ac}(k)= T \(( \ln{F(k,T)} - \frac{1}{q} \sum_{l=0}^{q-1} \ln{F(l,T)} \)),~~\tx{with}\\
F(k,T)= \sum_{l=0}^{q-1} \exp{
\Big (
\big (
J_{ab}(l)+J_{bc}
\(( \tx{mod} \(( \((q+k-l \)),q \)) \)) 
\big )
/T  
\Big )}.
\label{decimation}
\eea
To do a complete MKRG step, one has to pick up $2^{d-1}$ links decimated according to~(\ref{decimation}), to create $2^{d-2}$ new links and merge them into the final new $J$, this final step being just a link addition. We thus have to generate a set of $J_{ij}^n(k)$ at each renormalization step $n$; since we are dealing with a random system, we will do that numerically using the {\it pool} method of~\cite{BanavarBray87}, the only tricky points being to take care of the numerical representation of infinitesimal and exponential numbers, and to check that the pool size is not too small. One can use any specific form for the initials conditions $J^0_{ij}(\phi)$; we will choose here the random gauge XY spin glass, thus starting with $J^0_{ij}=\cos{\((\phi_i-\phi_j-A_{ij}  \))}$, where $A_{ij}$ are randomly chosen between $0$ and $2\pi$. Of course the initial cosine distribution is {\it not} conserved after iterations and converges to a random distribution; this is indeed happening for almost any initial distribution, leading to universality. An exception to this rule arises when the interaction has a reflexion symetry, like for instance $J_{ij} \vec{S_i}\vec{S_j}$; in that case there is no finite T transition for $d=3$, but any weak random perturbation changes the RG flow to that of the gauge model (see~\cite{CieplakBanavar92,CieplakLi93} for discussions on these points) which seems, therefore to be the more natural model to study.

\section{Energy, entropy, distance and chaos} For a complete description of method and formalism that will be used in the following, especially concerning the pool method, we refer the reader to~\cite{BanavarBray87}. The idea is to generate a representative set of $J^n$ at each iteration step. It is then possible to compute all quantities of interest. The droplet free energy and entropy reads
\be
\frac{1}{2} F(T) = 
{\la
\(( 
J^n(T)
\))^2
\ra
^{1/2}}
~~~~~~
\frac{1}{2}S(T) =\lim_{\delta T \rightarrow 0}
\frac{
\la 
\((J^n(T)-J^n(T+\delta T) \))^2
\ra 
^{1/2}
}
{\delta T}.
\ee
With q-states, one can choose to compute these quantities for the maximum $J(\phi)$ or to do an angle average; results are not significantly modified. The very standard way of studying temperature chaos is then to consider the length scale at which two replicas of the same system at temperature $T$ and $T+\delta T$ decorrelate; we will thus define  $\ell_c$ as the length $2^{n_0}$ at which $d\((n_0\)) = 1/e$, where the distance $d\((n\))$ is defined by:
\be
d^2\((n\)) = \frac{
\la
\((J^n(T)-J^n(T+\delta T)\))^2
\ra
}
{
\la
\((J^n(T)\))^2
\ra
+
\la
\((J^n(T+\Delta T)\))^2
\ra
}
\label{dist}
\ee
When $\delta T << T$ this length scale is known to be well estimated by eq.(\ref{lstar}) (see for example~\cite{BanavarBray87,AspelmeierBray02}), however we do not really care about the preciseness of this particular equation and therefore formula (\ref{dist}) will be used in the following to define $\ell_c$. We will now show results obtained from the renormalization process, first for the Ising then for the XY model.

\section{Ising spin glasses and the non universality of $\ell_c$}

In a recent and very interesting paper~\cite{AspelmeierBray02} $\ell_c$ was computed in $3d$ Ising spin glasses, first using the MKRG approach, then directly on a $3d$ lattice by considering interfacial energy and entropy and using eq.(\ref{lstar}). The authors conclude that $\ell_c$ is so big that it will be hard or almost impossible to see it in numerical simulations. Let us re-examine some of the aspects of chaos in the Ising case. The $\theta$ exponent is well known for Ising MKRG, and even analytical studies are possible~\cite{BouchaudKrzakala02} giving, for $d=3$, $\theta \approx 0.26$. The surface fractal dimension $d_s$ is also know to be $d-1$ on such MK lattices. We have considered variations of the chaos intensity when one uses $\pm J$, Gaussian and also a power-law coupling distribution (decreasing like $x^{1+0.5}$), and rescaled all energy distributions in order to obtain $T_c=1$. We worked at $T=0.1$ and computed droplet energies and entropies, showing as expected that $\theta$, as  well as the entropy exponent, are the same for all models. The important point (see fig.\ref{fig_Ising}) is that, as tails increase in the initial distribution, the energy stiffness decreases and the entropy increases. We thus expect that chaos increases with tails, and, computing distances, we see that this is indeed the case. Thus, in complete agreement with eq.(\ref{lstar}), we find that chaos increases when the energy decreases and the entropy increases. More tails mean that fluctuations are stronger so, in other words, chaos increases with {\it fluctuations} in the system: the more fluctuations there are, the more chaos there is. 

To conclude this preliminary work, it is worth making the following comments:(a) although it may seem a bit trivial, we see here that $\ell_c$ {\it is not a universal quantity} and indeed can change from more than a factor $10$ even in simple $3d$ Ising model; it is thus somehow a bit strange to compare experiments and simulations in such a precise way as to consider exact values of length scales to rule out or confirm scenarios.(b) Another important practical information is that, when one wants to find chaos in a system, or when doing dynamical simulations to see rejuvenation, {\it Gaussian couplings should be used instead of discrete $\pm J$}; in dynamical studies especially, where the coherence length growth is very slow, the difference predicted here should be very important.(c) Looking at the inset of fig.\ref{fig_Ising}, one should also notice that the chaotic length is not really well defined in the sense that there is no sharp transition between $\ell < \ell_c$ and $\ell > \ell_c$. Decorrelation is in fact continuous and there are also some changes for $\ell<\ell_c$. This is a crucial point (see for instance~\cite{JonssonYoshino02b,BerthierViasnoff02}) since it explains why rejuvenation can appear before the equilibrium length scale reaches $\ell_c$, as shown explicitly in~\cite{SasakiMartin02b,Sche}.

\section{XY spin glasses and the importance of chaos} 
\begin{figure}
\includegraphics[width=0.48\textwidth]{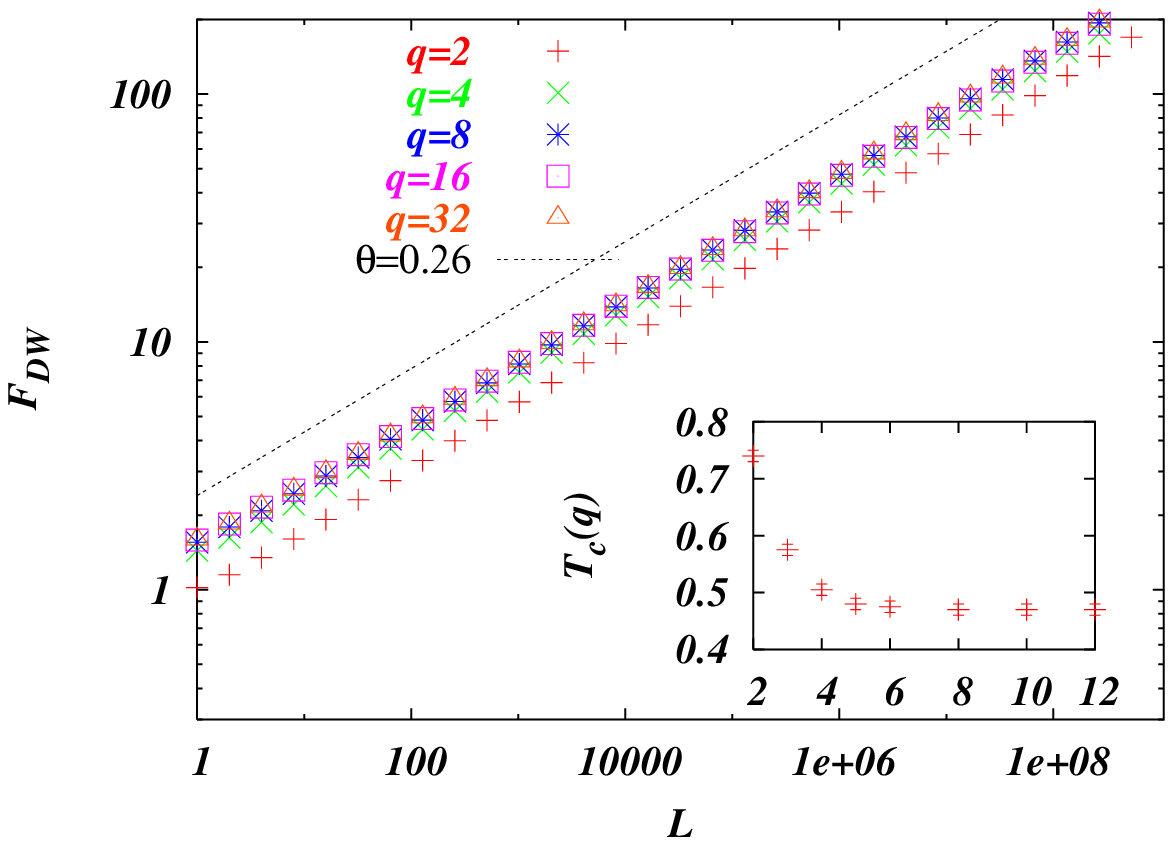}
\includegraphics[width=0.48\textwidth]{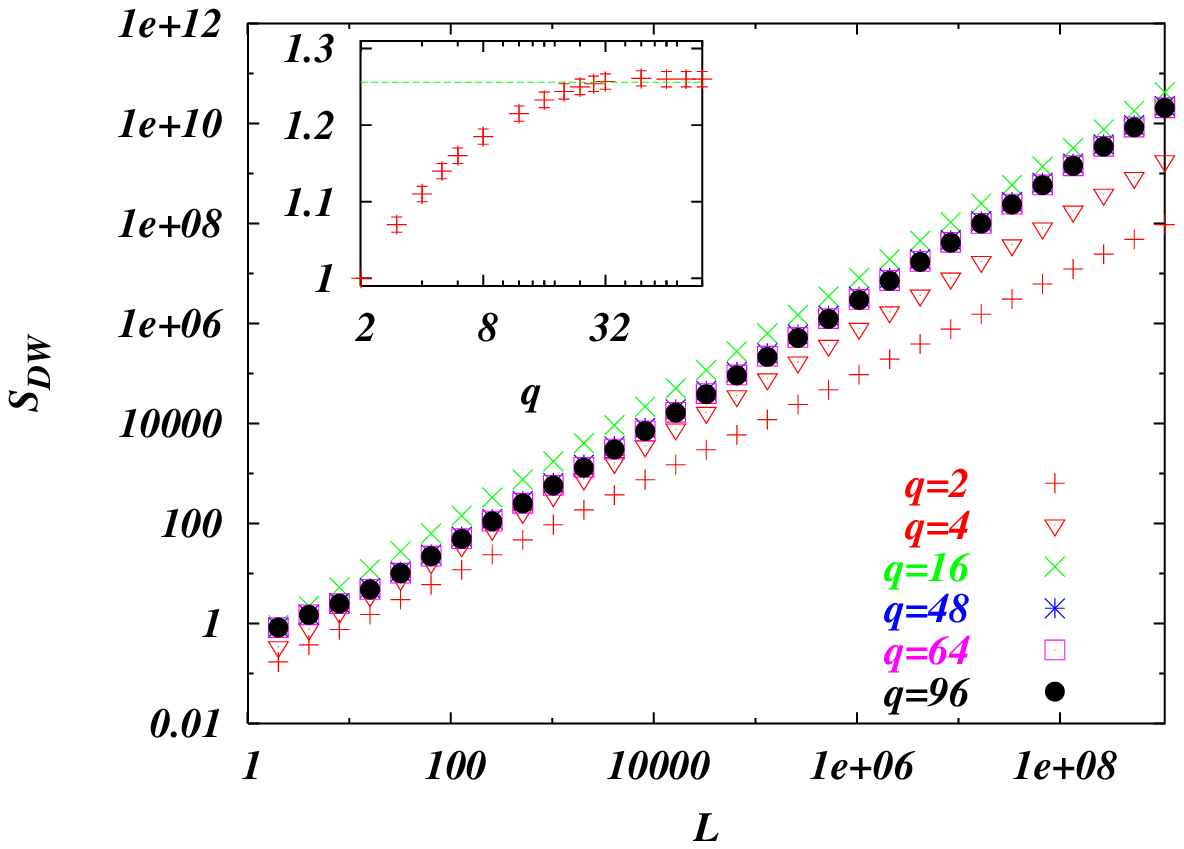}
\caption{Left: domain wall free energy of the q-states model at $T=0.01$ for different $q$ and the $0.26$ slope. In inset $T_c(q)$ converges fastly to $\approx 0.47$. Right: interfacial entropy at $T=0.01$, notice the non monotonic effect; convergence is reached for $q>40$ . In the inset the entropy exponent converges to $\approx 1.26$ (entropy was computed using $\delta T = 10^{-11}$).}
\label{fig_Tc}
\label{fig_Entro}
\end{figure}
We turn now to the q-states model and its XY limit. We will use the random gauge spin glass which, as opposed to the usual XY spin glass, has both its Ising ($q=2$) and XY ($q=\infty$) limits that display similar ordering with a finite $T_c$ on such lattices~\cite{CieplakBanavar92,CieplakLi93}. We thus start from a set of initial random link variables
\be
J_{ij}(k)=\cos{\(( 2 \pi k/q - A_{ij}\))}. 
\ee
A very nice study of chaotic properties of this model can be found in~\cite{CieplakLi93}. We will try here to be more precise, focusing on the values of $\ell_c$ at the $q \rightarrow \infty$ limit.
Let us first see how do interfacial entropy and free energy change with $q$. A first very important property is that the droplet exponent $\theta \approx 0.26$ is the same for all $q$ (see fig.\ref{fig_Tc}) suggesting that $T_c$ is always finite; we have indeed checked that $T_c$ remains finite at large $q$. We can thus compare, changing from Ising to XY, {\it similar} kinds of system with a {\it similar} low temperature droplet spin glass phase; this will allow us to isolate the effect of continuous spins on the chaotic length. 

If the free energy has the same $\theta$ exponent, the interfacial entropy is found to be {\it non} universal. However, the way the limit $q \rightarrow \infty$ is reached is a bit tricky due to a non monotonic effect: entropy first increases, then decreases and saturates (see fig.\ref{fig_Entro}). This effect is less strange after the following remark: if one starts with a random distribution of $J(\phi)$ using a uniform (instead of cosine) distribution  then {\it the effect disappears}; so this non monotonic effect comes from a pre-asymptotic regime, where the system is loosing its initial cosine coupling distribution. We need therefore to go to very large $q$ (here $q>40$) to see the XY regime. In contrast the entropy exponent {\it grows} monotonically from $d_s/2=1$ to $\approx 1.26$, in agreement with the result $\xi \approx 1$ for the chaos exponent when $q \rightarrow \infty$~\cite{CieplakLi93}. Finally, as expected, the interfacial entropy is bigger for XY spins and we can expect that so is chaos.

\begin{figure}
\includegraphics[width=0.5\textwidth]{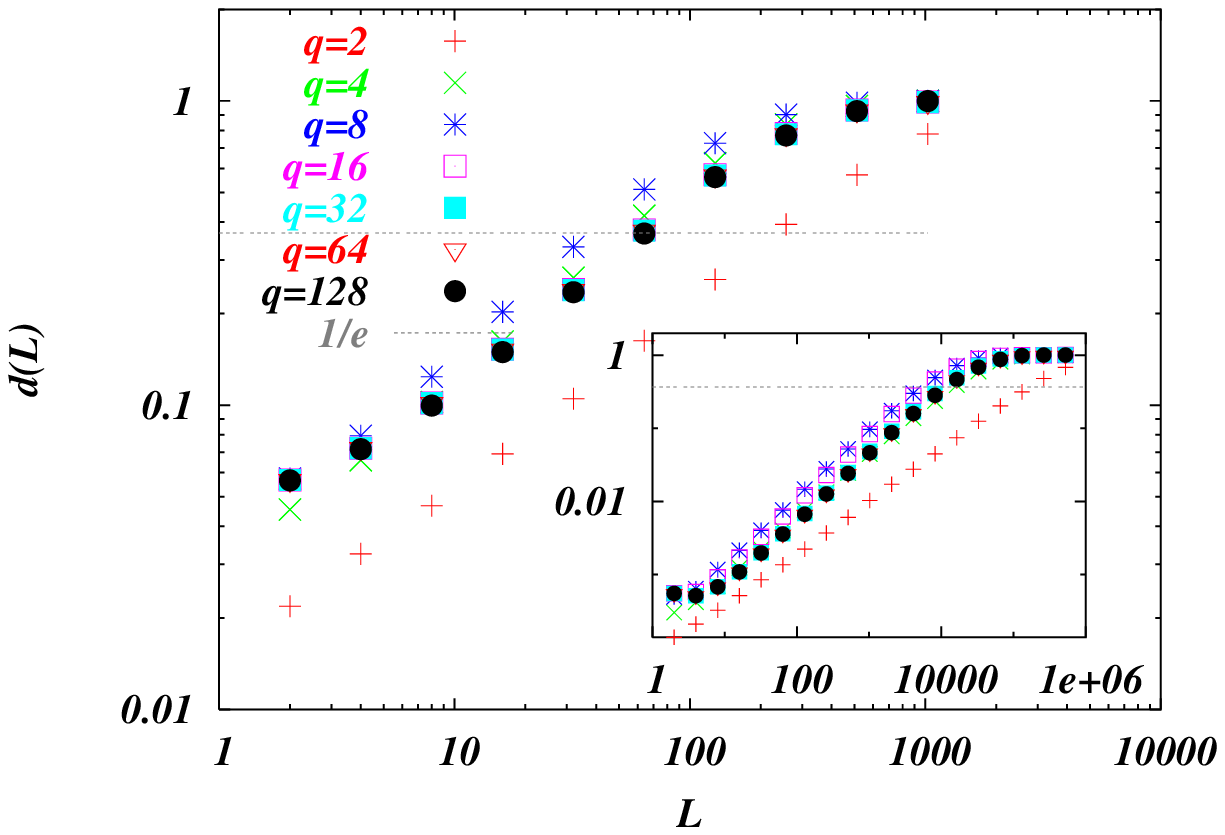}
\includegraphics[width=0.5\textwidth]{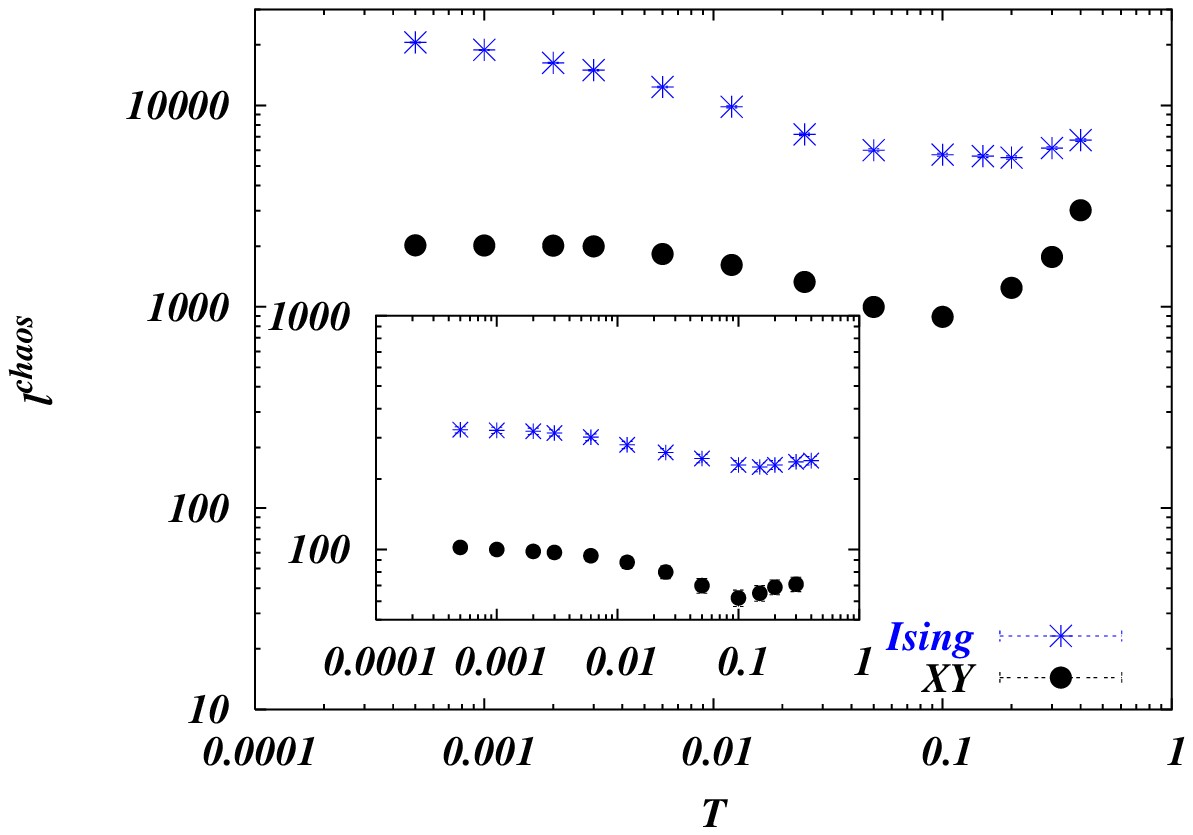}
\caption{Left: distance (see eq.~(\ref{dist})) between equilibrium configurations at temperature $T_1=0.1$ and  $T_2=0.2$ for different $q$. Symbols for $q=16,32,64$ are hidden by those for $q=128$, showing the convergence to the high $q$ limit. In the inset the same with $T_1=0.05$ and  $T_2=0.06$. Right: chaotic length $\ell_c$ for the XY model and Ising models. Note that $\ell_c$ can be about $10$ times smaller for XY than for Ising spins. This plot was made using $\delta T=0.01$ ($\delta T=0.1$ in the inset).}
\label{fig_chaos}
\label{fig_l}
\end{figure}

Now how does all that indeed translate in terms of the chaotic length? In~\cite{CieplakLi93} only the chaos exponent for $\delta T \rightarrow 0$ was considered and to go beyond that we need to compute $\ell_c$. In fig.\ref{fig_chaos} we plot the distance for different $q$, showing that we have again the non-monotonic effect: the distance first decreases, re-increases, and finally saturates with $q$ so that great care has to be taken for the limit $q \rightarrow \infty$.
That said, our final results clearly show that $\ell_c$ is smaller for XY than for Ising spins (the effect would have been even stronger starting from a uniform distribution, thus supressing the pre-asymptotic regime) and that it can decrease by {\it more than one decade}. This is illustrated in fig.\ref{fig_l} where we plot, for many temperatures in the Ising and the XY limit case, $\ell_c$ for $\delta T =0.1$ and $\delta T =0.01$ (the XY limit was obtained with $q=64$ for $\delta T=0.1$ and $q=128$ for $\delta T=0.01$). The conclusion is thus that, within this approach, the XY is more chaotic than the Ising spin glass, as expected from our previous argument.

\section{Discussion} Our main message is that, in a droplet/scaling approach, temperature chaos is far stronger for continuous than for Ising spins. Since such a chaotic temperature dependence {\it does} give rise to both {\it rejuvenation} and {\it memory}, as shown in ad-hoc models~\cite{YoshinoLemaitre00,JonssonYoshino03} as well as in dynamical MKRG approaches~\cite{SasakiMartin02b,Sche}, one can thus use it to interpret experiments~\cite{JonssonYoshino03}. That chaos is stronger for continuous spins then may explain naturally why rejuvenation effects are also experimentally stronger for continuous spins, a result sometimes presented as a riddle. Of course, since any simple picture with a growing length scale and changes in equilibrium configurations with temperature are sufficient to account for that kind of experiments~\cite{BerthierViasnoff02}, there are other possible explanations for these effects~\cite{BerthierBouchaud02,BerthierHoldsworth02}, but it is fair to say that they do not tell anything about differences between Ising and continuous spins; this is therefore an indication that chaos does play a role in these memory/rejuvanation protocols.

Note however, the very nature of the spin glass phase still being unknown and object of many debates, that all these scaling considerations might turn out too na\"{\i}ve for real systems. We considered in this letter only the simplest possible model, but we believe nevertheless that, if temperature chaos does exist, its generic behavior should be well described by this approach. Another potential problem concerns the nature of the spin glass phase with XY or Heisenberg spins, which could be different from the Ising one (like the chiral ordering of~\cite{Kawamura92}, however jeopardized in most recent numerical work~\cite{Young}).

Finally, our results also suggest some natural extensions:(i) Monte Carlo studies should be able to see chaos more easily in XY and Heisenberg than in Ising spin glasses, (ii) one may also play with the couplings to have a shorter $\ell_c$, since most work is done with $\pm J$ and that our results suggest instead to use a Gaussian or even a more tailed distribution, and (iii) dynamical studies of MK Ising spin glasses~\cite{SasakiMartin02b,Sche} may be usefully extended to XY spins.

\acknowledgments I would like to thank J-P.~Bouchaud, T.~J\"org, O.~C.~Martin and F.~Ricci-Tersenghi for their useful comments. I also benefit from discussions with colleagues at the 2003 SPHINX general meeting in Capoboi, Sardinia. I acknowledge support from European Community's Human Potential program under contract HPRN-CT-2002-00319 (STIPCO).

\bibliographystyle{prsty}
\bibliography{../../Bib/references}

\end{document}